\documentclass[aps,prd,onecolumn,preprintnumbers,nofootinbib, 12pt]{revtex4-1}
\usepackage[margin=1in]{geometry}

    \usepackage{amssymb,amsmath,amsthm,mathtools,physics}
\usepackage{slashed}
\usepackage{float}

\usepackage[linktoc=page]{hyperref}
\usepackage{cleveref}

\usepackage[usenames,dvipsnames]{xcolor}
\usepackage{subcaption}
\usepackage{graphicx}
\graphicspath{{./fig}}

\def\bar{\overline}

\def\vec{\vb*}

\def\d{\partial}

\DeclareMathOperator{\diag}{diag}
\def\ve{\varepsilon}

\def\chf11#1{{}_{1}F_{1}\left( #1 \right)}

\def\II{\mathbb{I}}

\def\cL{\mathcal{L}}

\def\cN{\mathcal{N}}

\def\cR{\mathcal{R}}

\begin{document}

\preprint{UCI-TR-2026-06}

\title{Revisiting fermion bound states in baby-Skyrme background\\ with Dzyaloshinskii--Moriya interaction}

\author{Arvind Rajaraman}
\affiliation{Department of Physics and Astronomy, 
University of California, Irvine, CA 92697-4575, USA
}

\author{Chao-Hsiang Sheu}
\affiliation{Department of Physics and Astronomy, 
University of California, Irvine, CA 92697-4575, USA
}

\author{Alexander Stewart}
\affiliation{Department of Physics and Astronomy, 
University of California, Irvine, CA 92697-4575, USA
}

\begin{abstract}

In this paper, we investigate a fermion coupled to a Skyrme model in $2+1$ dimensions, where the Skyrmion is stabilized by the Dzyaloshinskii-Moriya and Skyrme interactions under a quadratic potential. This framework interpolates between the magnetic Skyrmion at the critical coupling and the baby-Skyrme limit. 
The Dirac equation is studied both analytically in a non-relativistic reduction and numerically for the full relativistic spectrum, and the parameter region admitting states bound to the Skyrmion is determined. Localized solutions exist only for electrically charged fermions with positive charge and negative angular momentum, and are absent for neutral fermions. 
The lowest bound state in each angular momentum sector is characterized as a function of the fermion mass, charge, and the coupling $h$ to the Skyrmion isospin, and its behavior is compared across the magnetic, baby, and mixed Skyrmion backgrounds. 
The resulting fermion--Skyrmion composite constitutes an electrically charged state bound to a topological texture, providing concrete signatures for potential future transport and scattering measurements in chiral magnets.

\end{abstract}

\maketitle

\section{Introduction \label{sec:intro}}

Solitonic excitations, encompassing both topological and non-topological natures, are fundamental to our understanding of a variety of physical systems ranging from high-energy physics \cite{Rajaraman:1982is,Manton:2004tk,Shnir:2018yzp,Shifman:2022shi,Shifman_Yung_2023}, astrophysics and cosmology to condensed matter physics \cite{2002dgcm.book.....N}.
In high-energy physics, prominent examples include hypothetical magnetic monopoles predicted by Grand Unified Theories, as well as non-topological Q-balls and boson stars with various cosmological implications. In condensed matter physics, these excitations are essential for describing phenomena such as vortices and Hopfions in superconductors and the widely-studied magnetic Skyrmions.

Skyrmions in chiral magnets have attracted considerable attention in recent years due to their potential for next-generation technologies. 
The Skyrme model in $3+1$ dimensions \cite{Skyrme:1961vq,Skyrme:1962vh} is a nonlinear field theory of pions and was originally studied as a qualitative approximation of the baryonic sector of quantum chromodynamics (QCD). (See e.g. the reviews \cite{Manton:2004tk,Shifman:2022shi} and the references therein.) 
In two spatial dimensions, the similar Skyrmion systems have attracted significant recent attention due to their importance for magnetic storage devices \cite{2013NatNa...8..152F,Nagaosa:2013ftn,fert2017magnetic}, driven by the rapid development and escalating demands of memory device technologies in the artificial intelligence era.
The manipulation of skyrmionic excitations in memory devices is achieved via electric currents, which fundamentally motivates a thorough study of their interaction with fermions. 

 The form of the solitonic Skyrmion solutions stabilized by the Skyrme and Dzyaloshinskii--Moriya (DM) interactions has been investigated extensively~\cite{Hanada:2023lnm,Bolognesi:2024mjs}, as has their organization into Skyrmion lattices \cite{PhysRevB.91.224407,2006Natur.442..797R,PhysRevB.82.094429,2009Sci...323..915M,Yu:2010ayi,2011NatPh...7..713H,Ross:2020hsw,Amari:2024jxx}, see Ref. \cite{Gobel:2020mqd} for a recent review. Analogous constructions in lower dimensions have also been studied analytically \cite{Hongo:2019nfr,Ross:2020orc}, and the magnetic properties of these configurations, including their associated Hall responses, have been examined in detail~\cite{2017NatPh..13..112C,2011PhRvL.107m6804Z,2017NatPh..13..162J,2017NatPh..13..170L,PhysRevB.47.16419,PhysRevLett.102.186602,2015NatCo...6.8981B,Cook:2023jhu}. In contrast, the fermion--Skyrmion system has received much less attention. Existing studies have typically focused on a single stabilizing mechanism, considering either a baby-Skyrme interaction~\cite{Perapechka:2018yux,Barsanti:2021vhd} or a DM interaction~\cite{Perapechka:2019upv}, with magnetic effects incorporated through a Zeeman potential. By analyzing the fermion spectrum in such Skyrmion backgrounds, these works identified certain localized fermion solutions.

In this work, we consider the fermion--Skyrmion system in $2+1$ dimensions in a unified setting that simultaneously incorporates the DM and Skyrme stabilizing terms together with both the Zeeman and squared-Zeeman couplings. This setup reduces, in a particular limit, to the analytically tractable critical-coupling profile of Ref.~\cite{Barton-Singer:2018dlh}, while smoothly connecting the magnetic and baby-Skyrmion regimes within a single framework.

Our key findings are as follows.
Taking the fermion spectrum in the absence of the Skyrmion background as the reference $\varepsilon_{\mathrm{min}}$, so that a state is bound whenever $\varepsilon-\varepsilon_{\mathrm{min}}<0$, a variational argument in the non-relativistic limit shows that a bound state can form only when $|e|(|l-1|-|l|)+e>0$. This singles out the sector $e>0$, $l\leq 0$ and excludes the entire $l\geq 1$ tower. 
A neutral fermion ($e=0$) admits no bound state, since without the coupling to the external field no localized mode lies below the reference energy. 
Direct diagonalization of the full Dirac Hamiltonian confirms our analysis. Namely, we find no bound states outside $e>0$, $l\leq 0$, and within this sector exactly one bound state per angular momentum. Also, it persists numerically up to $h/m\sim\mathcal{O}(1)$ provided $h<m$, so the obstruction is robust beyond the strict non-relativistic expansion. 
The binding energy strengthens monotonically with the coupling $h$ and weakens as $|l|$ grows, the latter reflecting the centrifugal barrier in the effective potential and being well described by the shifted power law $E_{\mathrm{bind}}=-A/(|l|+c)^{\alpha}$.

Varying the ratio $\lambda/\kappa$ at fixed $\mu$ interpolates the background between the magnetic Skyrmion at the critical coupling $(\kappa,\lambda)=(1,0)$ and the baby Skyrmion $(\kappa,\lambda)=(0,1)$, and the binding energy is found to depend monotonically on this ratio. It is deepest for the magnetic Skyrmion, shallowest for the baby Skyrmion, and takes intermediate values for the mixed backgrounds. This ordering is captured by modeling the effective potential as a finite well of depth $V_0$ and width $a$ set by the Skyrmion profile. 
The magnetic Skyrmion has the larger spatial extent, so its broader well suppresses the centrifugal contribution $g(l^2)/a^2$ and yields deeper binding, whereas the more localized baby-Skyrmion profile enhances it.

The paper is organized as follows. In \cref{sec:model} we introduce the Skyrme model under consideration and review the relevant Skyrmion background. \Cref{sec:FSsystem} formulates the coupling of the Dirac fermion to this background and derives the radial eigenvalue problem governing the bound-state spectrum. The ground state energy far away from a Skyrmion is carried out for future comparison in \cref{sec:ENoSkyrm}. The non-relativistic reduction of this problem is carried out in \cref{sec:NR_limit}, where we establish the analytic criterion for the existence of states bound to the certain Skyrmion background. \Cref{sec:num} presents our numerical results based on the full Dirac equation, which confirm the non-relativistic prediction and extend the analysis to the relativistic regime. 

\section{Overview \label{sec:model}}

In this section, we begin by reviewing the Skyrmion ansatz and then proceed to the general analysis of the combined system.
We focus on the stationary configurations, i.e. there is no time dependence in both the Skyrmion and fermions.

The Skyrmion part of the model can be described by the following Lagrangian
\begin{align}\label{eq:Lskym}
\cL_{\rm Skyrm} =
\frac{1}{2} \d_{a}\vec{S} \cdot \d^{a}\vec{S}
- \kappa \vec{S} \cdot \left( \nabla \times \vec{S} \right)
- \frac{\lambda}{4} \left( \d_{a}\vec{S} \times \d_{b} \vec{S} \right) \cdot \left( \d^{a}\vec{S} \times \d^{b} \vec{S} \right)
- \frac{\mu^2}{2}\left( 1-S_{3} \right)^2 
\end{align}
where $a,b=0,1,2$ and $\vec{S}$ is a unit magnetization vector on a two-sphere.
In other words, the model includes the Dzyaloshinskii-Moriya interaction parametrized by the coupling constant $\kappa$, the Skyrme term with the coefficient $\lambda$.
The Minkowski signature $\eta_{\mu\nu} = \diag (1,-1,-1)$ is chosen throughout this paper.

The last term is the potential associated to the squared Zeeman interaction. 
One can rewrite the potential as
\begin{align}
    \frac{\mu^2}{2}(1-S_3)^2 = \mu^2(1-S_3) - \frac{\mu^2}{2}(1-S_3^2)
\end{align}
indicating the background magnetic field is $\vec{B} = \mu^{2} \hat{\vec{z}}$. For the coupling to fermions,
we will take the symmetric gauge for the corresponding gauge potential, i.e.
\begin{align}\label{eq:gaugef}
    A_{\nu} = \frac{\mu^{2}}{2} \bigl( 0,\, -y,\, x\bigr)
\end{align}
We can fix $\mu=1$ by a rescaling of the parameters.

Following the notation in \cite{Barton-Singer:2018dlh}, we consider the ansatz of the Skyrmion profile 
\begin{align}\label{eq:santz}
\vec{S}
= \Bigl(\sin{f(r)}\cos(\theta+\delta)\,, ~ \sin{f(r)}\sin(\theta+\delta) \,,~ \cos{f(r)}\Bigr)^T
\end{align}
subject to the boundary condition 
\begin{align}\label{eq:mskbc}
f(0) = \pi \qand f(\infty) = 0
\,.
\end{align}
The underlying real space is parametrized by the polar coordinate $(r,\theta)$ and $\delta$ represents the helicity of the Skyrmion. The helicity will be taken to be $\pi/2$ in the subsequent discussions as the background profile corresponds to a magnetic Skyrmion of the Bloch type\footnote{For magnetic Skyrmion, two kinds of Skyrmion configurations are often taken into account, say, the N\'eel type and the Bloch type. See \cite{Nagaosa:2013ftn,fert2017magnetic,Gobel:2020mqd} and references therein for visualizations and more detailed discussions. In our convention, the N\'eel type Skyrmion can be obtained by taking $\delta=0$ and promoting $\nabla$ to $\cR(\pi/2)\nabla$ where $\cR(\varphi)$ is the two dimensional rotational matrix.} in the $\lambda \to 0$ limit.
 
One ineteresting point is the  magnetic Skyrmion at critical coupling. Here, the Skyrme interaction is set to zero ($\lambda=0$), while the DM coefficient is tuned to match the potential parameter, specifically $\kappa = \mu = 1$.
The equation of motion of the Skyrmion field reduces to a first order equation and has an exact solution
\cite{Barton-Singer:2018dlh} 
\begin{align}
    f(r) = 2\arctan(\frac{2\kappa}{\mu^{2}r}) 
\end{align}

\begin{figure}[t]
    \centering
    \includegraphics[width=0.8\linewidth]{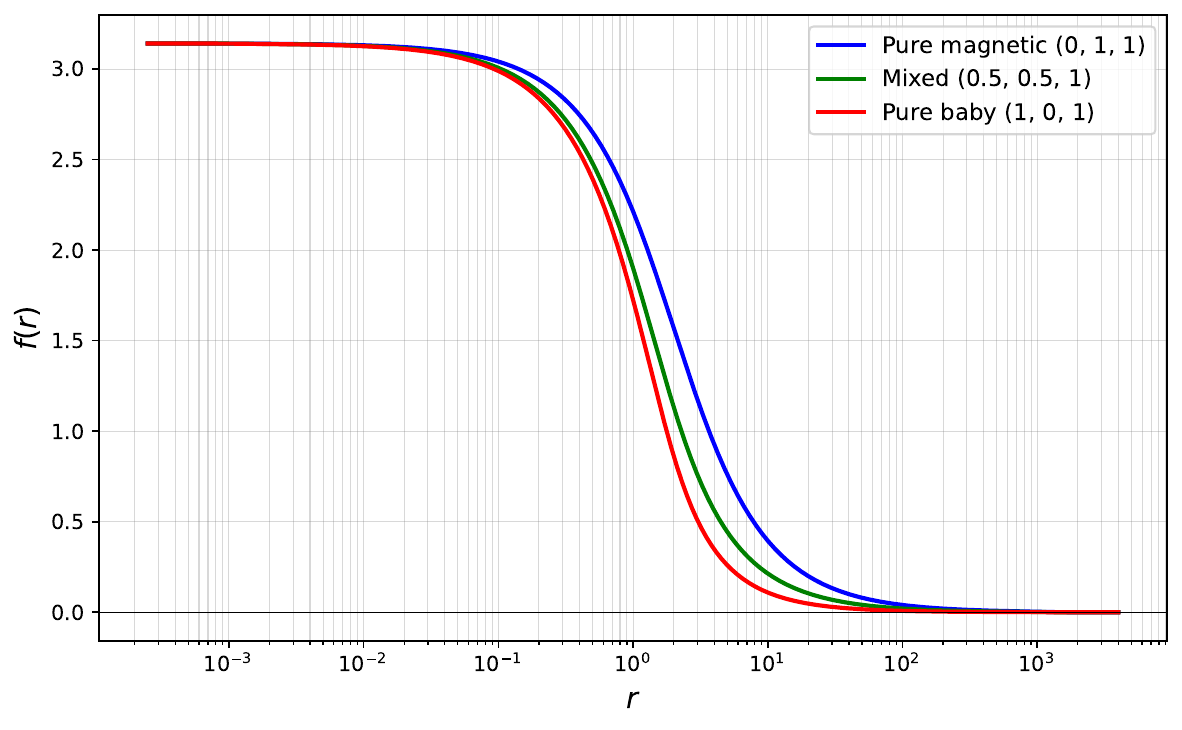}
    \vskip-3ex
    \caption{\small Skyrmion profiles without fermion backreaction for $(\kappa,\lambda,\mu) = (0,1,1)$, $(0.5,0.5,1)$, and $(1,0,1)$.}
    \label{fig:skypflh0}
\end{figure}

Other parameter points must be treated numerically using e.g. the boundary value solver in the \texttt{SciPy} package in Python.
In  \cref{fig:skypflh0}, we present the skyrmion profile  for three points  corresponding to $(\kappa,\lambda) = (1,0), (\kappa,\lambda) =(0,1)$, and $(\kappa,\lambda) =(0.5,0.5)$. 

\section{Fermion-Skyrmion system}
\label{sec:FSsystem}

We now consider  fermions in in this Skyrmion background.

The  spin-1/2 fermion $\Psi$ is a doublet under isospin. We will combine the spin and isospin components to a four-spinor notation
\begin{align}
\Psi=\Psi_{spin}\times \Psi_{isospin}=\left(\begin{array}{c}\Psi_{++}\\
\Psi_{+-}\\
\Psi_{-+}\\
\Psi_{--}
\end{array}\right)
\end{align}

The fermion Lagrangian takes the form
\begin{align}
\cL_{\rm f} &= 
\bar{\psi}(i {\gamma}^{\mu}D_{\mu}-m)\psi + h \vec{S} \cdot (\bar{\psi}\vec{\tau}\psi)
-h  \cdot (\bar{\psi}\vec{\tau}_3\psi)
\end{align}
where $D_{\mu}$ denotes the covariant derivative defined as $D_{\mu} := \d_{\mu} - ieA_{\mu}$ with the background gauge field given in \cref{eq:gaugef}. ${\gamma}^{\mu}$ and $\vec{\tau}$ represent spin and isospin matrices, respectively,
\begin{align}
{\gamma}^{0}  =\left(\begin{array}{cc}\II_2 &0\\
0&-\II_2 
\end{array}\right) \,, \quad   {\gamma}^{1}  =\left(\begin{array}{cc}
0 & -i\II_2\\
-i\II_2 & 0 
\end{array}\right) 
\,, \quad   {\gamma}^{2}  =\left(\begin{array}{cc}0 &-\II_2\\
\II_2&0
\end{array}\right)
\,, \quad   
{\vec{\tau}}  =\left(\begin{array}{cc}\vec{\sigma} &0\\
0& \vec{\sigma}
\end{array}\right) 
\end{align}

With the Skyrmion ansatz, the explicit matrix form of the Hamiltonian operator is 
\begin{align}\label{eq:fermHam}\small
H_{f} = {\mqty(
m-h \cos{f}
& -he^{-i(\theta+\delta)}\sin{f}
& -e^{-i\theta}\left( \d_{r} - \frac{i\d_{\theta}}{r} - \frac{e\mu^2 r}{2} \right)
& 0
\\[4mm]
-he^{i(\theta+\delta)}\sin{f}
& m+h \cos{f}
& 0
& -e^{-i\theta}\left( \d_{r} - \frac{i\d_{\theta}}{r} - \frac{e\mu^2 r}{2} \right)
\\[4mm]
e^{i\theta}\left( \d_{r} + \frac{i\d_{\theta}}{r} + \frac{e\mu^2 r}{2} \right)
& 0
&-m+h\cos{f}
& he^{-i(\theta+\delta)}\sin{f}
\\[4mm]
0
& e^{i\theta}\left( \d_{r} + \frac{i\d_{\theta}}{r} + \frac{e\mu^2 r}{2} \right)
&he^{i(\theta+\delta)}\sin{f}
&-m-h\cos{f}
)}
\end{align}
We will consider stationary solutions  $\psi(x_{\mu}) = e^{-i\ve t}\psi(x,y)$ .

In addition, one can simplify the analysis by noticing that the fermion Hamiltonian commutes with the angular momentum operator \cite{Perapechka:2018yux,Perapechka:2019upv,Amari:2019tgs,Loginov:2021rka,Barsanti:2021vhd,Loginov:2024nmi,Amari:2025rgt}
\begin{align}
\label{eq:angM}
K_3 := (\II_2\otimes\II_2)\left( -i\pdv{\theta} \right) + \frac{1}{2} \sigma_3\otimes\II_2 
+ \frac{1}{2} \tau_{3}
\end{align}
The eigenspinor of angular momentum can be written as 
\begin{align}\label{eq:fermastz}
    \psi_{l}(r,\theta) := \cN^{-1} \cdot \mqty(
        u_{1}(r)e^{i(l-1)\theta}\,, &
        iu_{2}(r)e^{il\theta}\,, &
        u_{3}(r)e^{il\theta}\,, &
        iu_{4}(r)e^{i(l+1)\theta} 
    )^{T}
\end{align}
where $l$ is the eigenvalue of $K_3$ acting on $\psi$. Since we are interested in the single fermion state, the normalization constant $\cN$ satisfies
\begin{align}
    1 = 2\pi\cN^{-2} \int_{0}^{\infty}\left( u_{1}^2+u_{2}^2+u_{3}^2+u_{4}^2 \right) r\dd{r}
\end{align}
where the expression in the integral is the fermion density 
\begin{align}
    \rho_{\rm ferm} = \psi^{\dagger}\psi
    = \cN^{-2}\Bigl( \abs{u_1}^2 + \abs{u_2}^2 + \abs{u_3}^2 + \abs{u_4}^2 \Bigr)
\end{align}
To trace the particle and the anti-particle states, we consider the quantity 
\begin{align}
Q := \int_{0}^{\infty} \left( u_1^2+u_2^2 - u_3^2 - u_4^2 \right) r\dd{r}
\end{align}
and will keep it positive in our numerical search.
The fermion eigenvalue equations then become
    \begin{align}\label{eq:bddbck}\begin{aligned}
        u_{1}' + \left( \frac{e\mu^2 r}{2} - \frac{l-1}{r} \right) u_{1} + \left(h \cos{f} -m \right)u_{3} + h\sin{f} \cdot u_{4} 
        &= \ve u_{3}
        \\[2mm]
        u_{2}' + \left( \frac{e\mu^2 r}{2} - \frac{l}{r} \right) u_{2} + h\sin{f} \cdot u_{3} - \left(h\cos{f} + m  \right)u_{4} 
        &= \ve u_{4}
        \\[2mm]
        -u_{3}' + \left( \frac{e\mu^2 r}{2} - \frac{l}{r} \right) u_{3} - \left( h\cos{f} -m \right) u_{1} - h\sin{f} \cdot u_{2} 
        &= \ve u_{1}
        \\[2mm]
        -u_{4}' + \left( \frac{e\mu^2 r}{2} - \frac{l+1}{r} \right) u_{4} - h\sin{f} \cdot u_{1} + \left(h \cos{f} + m  \right)u_{2} 
        &= \ve u_{2}
    \end{aligned}
    \end{align}


The boundary condition of the fermion eigenfunction is specified as 
\begin{align}\label{eq:fermBC}
    u_{i}(\infty) = 0
    \qand
    \left\{\begin{aligned}
        & u_{i}(0) = 0
        && \mbox{if $l_i \neq 0$}
        \\
        & u_{i}'(0) = 0
        && \mbox{if $l_i = 0$}
    \end{aligned}\right.
\end{align}
where $i=1,2,3,4$. Here $(l_1,l_2,l_3,l_4) = (l-1,\, l,\, l,\, l+1)$. The boundary condition around the origin can be seen from the regularity analysis around $r \to 0$ below.

It is instructive to examine the asymptotic behavior of the fermion wavefunction.
First, in the neighborhood of the origin, the regularity condition requires the fermion eigenfunction to be 
\begin{align}\label{eq:reg_cond}
    u_{i}(r) = r^{\abs{l_{i}}}  \left[ c_{i} + \order{r^{2}} \right]
\end{align}
The equation of motion can be used to reduce the  four parameters $c_{i}$ to two independent ones. For example, for $l=0$, 
\begin{align}
    c_1^{(l=0)} = \frac{\ve+h+m}{2}c_3^{(l=0)}
    \,,\quad
    c_2^{(l=0)} = -\frac{2}{\ve+h-m}c_4^{(l=0)}
    \,.
\end{align}

Near $r \to \infty$, the neutral $(e=0)$ fermion profile  can be approximated by the modified Bessel functions 
\begin{align}\label{eq:chglsasym}
    u_{i}(r) \sim K_{\abs{l_{i}}}\left( \sqrt{(h \pm m)^2-\ve^2} \cdot r \right)
\end{align}
decaying exponentially at large $r$. \Cref{eq:chglsasym} also indicates that 
\begin{align}
    \ve \leq \min(\abs{h+m},\abs{h-m}) 
\end{align}
for the bound state solutions.

On the other hand, the profile of a charged fermion\footnote{Note that in the decoupling limit with non-zero fermion charge, i.e. $h=0$ and $e \neq 0$, the system describes a charged fermion in a uniform magnetic field. Thus, the fermion energy spectrum is given by the Landau levels. This is also evident from the solution of \cref{eq:bddbck} in the present limit where the eigenstate in this limit turns out to be a generalized hypergeometric function ${}_{1}F_{1}(a,b,z)$ with the first parameter $a$ depending on $\ve$. The boundary condition at the infinity requires $a$ to be a non-positive integer, which quantizes the spectrum and reproduces the Landau levels.} scales as
\begin{align}
    u_{i}(r) \sim K_{0}\left( \frac{\abs{e}\mu^2}{4}r^{2} \right)
\end{align}
at large $r$. In the charged case, we would expect a more rapid decay when $r$ approaches the infinity by virtue of the $\exp(-\abs{e}\mu^2r^2/4)$ dependence.

Before turning to the main analysis, we briefly comment on the backreaction of the fermion on the Skyrmion background.
Although the fermion bilinear in principle sources the Skyrmion equation of motion, we have checked numerically that this effect produces only a minor distortion of the profile and leaves the bound-state spectrum essentially unchanged. We therefore treat the Skyrmion as a fixed background in this paper.

 \section{Energy In the absence of the Skyrmion}
 \label{sec:ENoSkyrm}

Before we take the non-relativistic limit, we will compute the ground state energy in the absence of the Skyrmion $(f(r) = 0)$ to compare to the ground state energy for a charged fermion $(e \neq 0)$ in the Skyrmion background. In \cref{eq:bddbck}, the $u_1$ and $u_3$ channel decouples from the $u_2$ and $u_4$ channel. We will focus on the $u_1$ and $u_3$ channel, the computation for the $u_2$ and $u_4$ channel is similar. We find
\begin{subequations}
\begin{equation}
    \varepsilon u_1 = (m-h)u_1+\bigg(-\frac{\d}{\d r}+\frac{e\mu^2 r}{2} - \frac{l}{r}\bigg)u_3\;,
\end{equation}
\begin{equation}
    \varepsilon u_3 = (h-m)u_3 + \bigg(\frac{\d}{\d r}+\frac{e\mu^2 r}{2}-\frac{l-1}{r}\bigg) u_1\;.
\end{equation}
\end{subequations}
Solving the latter equation for $u_3$, we find
\begin{equation}
    u_3 = \frac{1}{\varepsilon +m - h}\bigg(\frac{\d}{\d r}+\frac{e\mu^2 r}{2} - \frac{l-1}{r}\bigg)u_1\;.
\end{equation}
Using this to eliminate $u_3$ from the other expression, we arrive at
\begin{equation}
    \varepsilon^2 u_1 = (m-h)^2 u_1 +\bigg(-\frac{\d^2}{\d r^2} - \frac{1}{r}\frac{\d}{\d r} + \frac{(l-1)^2}{r^2}+\frac{e^2 \mu^4 r^2}{4} - e l \mu^2\bigg) u_1\;.
\end{equation}
The second term is the 2d radial Laplacian $-\nabla^2$ with angular quantum number $|l-1|$, along with a 2d radial harmonic oscillator with frequency $\omega = |e|\mu^2 /2$. The eigenvalues are labeled integer $n_1$ as
\begin{equation}
    \varepsilon = \sqrt{(m-h)^2 +|e|\mu^2(2n_1+|l-1|+1)-el\mu^2}\;.
\end{equation}
A similar analysis for the $u_2$ and $u_4$ channel leads to the energy eigenvalues
\begin{equation}
    \varepsilon = \sqrt{(m+h)^2 + |e|\mu^2(2n_2+|l|+1)-e(l+1)\mu^2}\;,
\end{equation}
with $n_2$ an integer as well. Generally, these two energy eigenvalues will not agree for arbitrary parameters and the consistent solutions have either $u_1 = u_3 = 0$ or $u_2 = u_4 = 0$. The ground state energy ($n_1=0$ and $n_2=0$) in the absence of the Skyrmion is then
\begin{equation}\label{eq:E_Landau}
    \varepsilon_{\mathrm{min}} = \min\bigg( \sqrt{(m-h)^2 + |e|\mu^2(|l-1|+1)-el\mu^2}\;,\; \sqrt{(m+h)^2 + |e|\mu^2(|l|+1)-e(l+1)\mu^2} \bigg)\;.
\end{equation}
If we can solve for the energy $\varepsilon$ of the fermion in the Skyrmion background, then the fermion is said to be bound to the Skyrmion whenever $\varepsilon < \varepsilon_{\mathrm{min}}$, with binding energy $E_{\mathrm{bind}} = \varepsilon - \varepsilon_{\mathrm{min}}$.

For the case of an uncharged fermion $(e=0)$, a similar analysis shows that the system in the absence of the Skyrmion reduces to two decoupled systems $(u_1,u_3)$ and $(u_2,u_4)$ of free fermions  with effective masses $m-h$ and $m+h$ respectivley. Since we focus on the case $h>0$, the minimum energy state for the uncharged fermion is always given by the $(u_1,u_3)$ channel with $\varepsilon_{\mathrm{min}} = m-h$.

\section{Non-relativistic limit}\label{sec:NR_limit}

Now we examine the fermion-Skyrmion system in the non-relativistic (NR) limit where $h \ll m$ and $\varepsilon = E + m$ with $|E| \ll m$. It is convenient for notational purposes to define the following quantities which appear repeatedly in computations
\begin{align}
    &\mathcal{D}_1 u_1 \equiv u_1' + \bigg(\frac{e\mu^2 r}{2}-\frac{l-1}{r}\bigg)u_1\;,\quad \mathcal{D}_2 u_2 \equiv u_2' + \bigg(\frac{e \mu^2 r}{2}-\frac{l}{r}\bigg)u_2\;,\\
    &m_{\pm} \equiv m\pm h\;,\quad E_\pm \equiv \varepsilon - m_{\pm}\;.
\end{align}
The $m_\pm$ and $E_\pm$ are the rest masses and NR energies of the two channels in the absence of the Skyrmion. Note that the $m_\pm$ and $E_\pm$ are not all independent, since $m_+ - m_- = E_+ - E_- = 2h$. Expanding the first two equations in \cref{eq:bddbck} to leading order in $E_-/m_-$ and $E_+/m_+$, we find
\begin{align}
    \begin{aligned}
        &u_3 = \frac{\big[2m_+-h\big(1-\cos f(r)\big)\big]\mathcal{D}_1 u_1 + h \sin f(r) \mathcal{D}_2 u_2}{4m_+m_- + 2h^2\big(1-\cos f(r)\big)}\;,\\
        &u_4 = \frac{\big[2m_- + h\big(1-\cos f(r)\big)\big]\mathcal{D}_2 u_2 + h \sin f(r) \mathcal{D}_1 u_1}{4m_+m_- + 2h^2\big(1-\cos f(r)\big)}\;.
    \end{aligned}
\end{align}
Note that in the absence of the Skyrmion background, these equations reduce to
\begin{equation}
    u_3 = \frac{\mathcal{D}_1 u_1}{2m_-}\;,\quad u_4 = \frac{\mathcal{D}_2 u_2}{2m_+}\;,
\end{equation}
which agrees with our interpretation of $m_\pm$ being the rest masses of the two channels in the absence of the Skyrmion. We can substitute these forms of $u_3$ and $u_4$ into the latter two equations of \cref{eq:bddbck} to eliminate $u_3$ and $u_4$ from the system, leaving two second order equations for $u_1$ and $u_2$. Up to  $\mathcal{O}(h/m_\pm)$ terms, the resulting second order equations can be written as
\begin{equation}\label{eq:NR_eqns}
    \varepsilon\begin{pmatrix}
        u_1 \\
        u_2
    \end{pmatrix} = H_{\mathrm{NR}}\begin{pmatrix}
        u_1 \\
        u_2
    \end{pmatrix} = \left[\begin{pmatrix}
        H_0^{(1)}+m_- & 0\\
        0 & H_0^{(2)}+m_+
    \end{pmatrix} + V_{\mathrm{Skyrm}(r)} \right]\begin{pmatrix}
        u_1 \\
        u_2
    \end{pmatrix}\;,
\end{equation}
where the $H^{(i)}_0$ denote\footnote{Here, $\nabla^2$ denotes the radial Laplacian $\nabla^2 = \partial_r^2 + \frac{1}{r}\partial_r$.}
\begin{align}\label{eq:H0_NR}
    \begin{aligned}
        &H_0^{(1)} = \frac{1}{2m_-}\bigg[ -\nabla^2 + \frac{(l-1)^2}{r^2} + \frac{e^2 \mu^4 r^2}{4} - el\mu^2\bigg]\;,\\
        &H_0^{(2)} = \frac{1}{2m_+}\bigg[ -\nabla^2 + \frac{l^2}{r^2} + \frac{e^2 \mu^4 r^2}{4} - e(l+1)\mu^2\bigg]\;,
    \end{aligned}
\end{align}
and $V_{\mathrm{Skyrm}}(r)$ denotes a localized potential generated by the Skyrmion-fermion couplings,
\begin{equation}
    V_{\mathrm{Skyrm}}(r) = \begin{pmatrix}
        h\big(1-\cos f) & -h \sin f\\
        -h \sin f & -h\big(1-\cos f\big)
    \end{pmatrix}\;.
\end{equation}
Notice that the potential $V_{\mathrm{Skyrm}}(r)$ vanishes as $r \rightarrow \infty$.

 

\subsection{Analytic results for the non-relativistic limit for $e\neq 0$ }\label{ssec:analytic_NR_e_neq_0}

Here we analyze the NR system in the case that $e \neq 0$ and drop corrections of order $\mathcal{O}(h/m)$ or $\mathcal{O}(|e|\mu^2/m^2)$. In the absence of the Skyrmion, the potential $V_{\mathrm{Skyrm}}(r)$ vanishes. As a result, the Hamiltonian becomes
\begin{equation}
    H_{\mathrm{NR}}\rightarrow  \begin{pmatrix}
        H_0^{(1)} + m_- & 0
        \\
        0 & H_0^{(2)} + m_+
    \end{pmatrix}\;,
\end{equation}
up to $\mathcal{O}(h/m)$ and $\mathcal{O}(|e|\mu^2/m^2)$ corrections. $u_1$ and $u_2$ have decoupled and the system just describes two NR Landau Hamiltonians. Their spectra are
\begin{align}
\begin{aligned}
        &\mathrm{spec}\big(H_0^{(1)}+m_-\big) = m_- + \frac{|e|\mu^2 (2n_1 + |l-1|+1) - el\mu^2}{2m_-}\;,\\
        &\mathrm{spec}\big(H_0^{(2)}+m_+\big) = m_+ + \frac{|e|\mu^2 (2n_2 + |l|+1) - e(l+1)\mu^2}{2m_+}\;,
\end{aligned}
\end{align}
with $n_1=0,1,2,\dots$ and $n_2=0,1,2,\dots$ both non-negative integers. The minimal energy states occur for $n_1=0$ and $n_2=0$, with
\begin{align}
\begin{aligned}
        &\min\mathrm{spec}\big(H_0^{(1)}+m_-\big) = m_- + \frac{|e|\mu^2(|l-1|+1)-el\mu^2}{2m_-}\equiv m_- + A_1\;,\\
        &\min\mathrm{spec}\big(H_0^{(2)}+m_+\big) = m_+ + \frac{|e|\mu^2(|l|+1)-e(l+1)\mu^2}{2m_+}\equiv m_+ + A_2\;.
\end{aligned}
\end{align}
All of these states are the standard Landau bound states, but we can ask if the presence of the Skyrmion increases or decreases the binding of the Landau levels. To check this, we look for states $\chi$ with energy $\varepsilon$ such that $\varepsilon < \varepsilon_{\mathrm{min,NR}}$ where
\begin{equation}
    \varepsilon_{\mathrm{min,NR}} = \min\big(m_- + A_1, m_+ + A_2\big) = m_- + \min\big(A_1,A_2 + 2h\big)\;.
\end{equation}

Consider a bound state $\chi = (u_1,u_2)^\mathrm{T}$ with energy $\varepsilon$, normalized such that $\braket{\chi} = 1$. Up to $\mathcal{O}(h/m)$ and $\mathcal{O}(|e|\mu^2/m^2)$ corrections, the Hamiltonian for the fermion in the presence of the Skyrmion is described by
\begin{equation}
    \varepsilon\begin{pmatrix}
        u_1\\
        u_2
    \end{pmatrix} = \begin{pmatrix}
        H_0^{(1)}+m_- + h\big(1-\cos f\big) & -h\sin f\\
        -h \sin f & H_0^{(2)}+m_+ - h\big(1-\cos f\big)
    \end{pmatrix}\begin{pmatrix}
        u_1\\
        u_2
    \end{pmatrix} = H\chi\;.
\end{equation}
We can rewrite the Hamiltonian here as
\begin{equation}
    H = (m_-)\,\boldsymbol{1}_2 + \begin{pmatrix}
        H_0^{(1)} & 0\\
        0 & H_0^{(2)}
    \end{pmatrix} + M\;,
\end{equation}
where
\begin{equation}
    M = \begin{pmatrix}
        0 & 0\\
        0 & 2h
    \end{pmatrix} + V_{\mathrm{Skyrm}}(r) = \begin{pmatrix}
        h\big(1 - \cos f\big) & -h\sin f\\
        -h\sin f & h\big(1+\cos f\big)
    \end{pmatrix}\;.
\end{equation}
The eigenvalues of $M$ are $2h$ and $0$, so necessarily we have that
\begin{equation}
    \bra{\chi}M\ket{\chi} \geq 0\;
\end{equation}
for normalizable state $\chi$. Combined with the bounds
\begin{equation}
    \bra{u_1}H_0^{(1)}\ket{u_1} \geq ||u_1||^2A_1\;,\quad \bra{u_2}H_0^{(2)}\ket{u_2} \geq ||u_2||^2A_2\;,
\end{equation}
we find
\begin{align}
    \varepsilon = \bra{\chi}H\ket{\chi} &= m_- + \bra{u_1}H_0^{(1)}\ket{u_1} + \bra{u_2}H_0^{(2)}\ket{u_2} + \bra{\chi}M\ket{\chi}\nonumber\\
    &\geq m_- + ||u_1||^2 A_1 + ||u_2||^2 A_2\nonumber\\
    &\geq m_- + \min\big(A_1,A_2\big)\;.
\end{align}

Now we compare this to $\varepsilon_{\mathrm{min,NR}}$,
\begin{align}
    \varepsilon < \varepsilon_{\mathrm{min,NR}} &\implies m_- + \min\big(A_1,A_2\big) < m_- +\min\big(A_1,A_2+2h\big)\nonumber\\
    &\implies \min\big(A_1,A_2\big) < \min\big(A_1,A_2+2h\big)\;.
\end{align}
Clearly, this inequality can only be satisfied if $A_2 < A_1$. From the explicit forms of $A_2$ and $A_1$, this reads
\begin{equation}
    \frac{|e|\mu^2(|l|+1)-e(l+1)\mu^2}{2m_+} < \frac{|e|\mu^2(|l-1|+1)-el\mu^2}{2m_-}\;.
\end{equation}
Rewriting this and dropping terms of order $\mathcal{O}(h/m)$, we recover
\begin{equation}
    e + |e|\big(|l-1|-|l|\big) >0\;.
\end{equation}
Note that for $l\geq 1$, $|l-1|-|l| = -1$, while for $l \leq 0$ we have $|l-1|-|l| = 1$. Consider first the case that $e < 0$, where the inequality becomes
\begin{equation}
    |l-1|-|l| > 1\;,
\end{equation}
which is not possible for any $l$. Now consider $e > 0$, where the inequality reads
\begin{equation}
    |l-1|-|l| > -1\;,
\end{equation}
which is possible for $l\leq 0$. Therefore, for the $e \neq 0$ case the NR analysis indicates there are bound states for $e > 0$, $l \leq 0$ while not providing evidence for bound states in any other sectors. Numerically, we do indeed find bound states only in the $e>0$, $l\leq 0$ sector.

\subsection{Analytic results for the non-relativistic limit for $e=0$}
For the $e=0$ case, the fermions do not experience any potential from the external magnetic field. In the absence of the Skyrmion, the potential $V_{\mathrm{Skyrm}}(r)$ again vanishes. As in the $e\neq 0$ case, the Hamiltonian becomes
\begin{equation}
    H\rightarrow \begin{pmatrix}
        H_0^{(1)} + m_- & 0
        \\
        0 & H_0^{(2)} + m_+
    \end{pmatrix}\;,
\end{equation}
up to $\mathcal{O}(h/m)$ corrections. The $H_0^{(i)}$ in the $e=0$ case are
\begin{align}
\begin{aligned}
        &H_0^{(1)} = \frac{1}{2m_-}\bigg[ -\nabla^2 + \frac{(l-1)^2}{r^2}\bigg]\;,\\
        &H_0^{(1)} = \frac{1}{2m_+}\bigg[ -\nabla^2 + \frac{l^2}{r^2}\bigg]\;.
\end{aligned}
\end{align}
In the absence of the Skyrmion, we see the $u_i$ describe free particles of masses $m_-$ and $m_+$. In contrast to the $e \neq 0$ case, there are no Landau bound states and instead there is just the continuum of scattering states. A bound state $\chi$ then needs an energy $\varepsilon$ such that $\varepsilon < \varepsilon_{\mathrm{min,NR}}$ where
\begin{equation}
    \varepsilon_{\mathrm{min,NR}} = \min (m_- , m_+) = m_-\;,
\end{equation}
since we study $h>0$. Following the same structural argument of the $e\neq 0$ case, we consider a bound state $\chi = (u_1, u_2)^\mathrm{T}$ with energy $\varepsilon$, normalized such that $\braket{\chi}{\chi} = 1$. Up to $\mathcal{O}(h/m)$ corrections, the Hamiltonian for the fermion in the presence of the Skyrmion is again described by
\begin{equation}
    \varepsilon\begin{pmatrix}
        u_1\\
        u_2
    \end{pmatrix} = \begin{pmatrix}
        H_0^{(1)}+m_- + h\big(1-\cos f\big) & -h\sin f\\
        -h \sin f & H_0^{(2)}+m_+ - h\big(1-\cos f\big)
    \end{pmatrix}\begin{pmatrix}
        u_1\\
        u_2
    \end{pmatrix} = H\chi\;.
\end{equation}
We can rewrite the Hamiltonian here as
\begin{equation}
    H = (m_-)\,\boldsymbol{1}_2 + \begin{pmatrix}
        H_0^{(1)} & 0\\
        0 & H_0^{(2)}
    \end{pmatrix} + M\;,
\end{equation}
where
\begin{equation}
    M = \begin{pmatrix}
        0 & 0\\
        0 & 2h
    \end{pmatrix} + V_{\mathrm{Skyrm}}(r) = \begin{pmatrix}
        h\big(1 - \cos f\big) & -h\sin f\\
        -h\sin f & h\big(1+\cos f\big)
    \end{pmatrix}\;.
\end{equation}
We can compute $\varepsilon$ as
\begin{equation}
    \varepsilon = \bra{\chi}H\ket{\chi} = m_- +\bra{u_1}H_0^{(1)}\ket{u_1}+\bra{u_2}H_0^{(2)}\ket{u_2} + \bra{\chi}M\ket{\chi}\;.
\end{equation}
We have that\footnote{This can be shown by integrating the second derivative term in $\nabla^2$ by parts, and using the regularity conditions on the $u_i$ at $r=0$ as well as their exponential decay as $r\rightarrow \infty$ to drop boundary terms and write the integrand in a manifestly positive form.} $\bra{u_1}H_0^{(1)}\ket{u_1} \geq 0$ and $\bra{u_2}H_0^{(2)}\ket{u_2} \geq 0$. $\bra{\chi}M\ket{\chi}$ is explicitly
\begin{align}
    \bra{\chi}M\ket{\chi} &= \int_0^{\infty}r\mathrm{d} r\;\left[\begin{pmatrix}
        u_1 & u_2
    \end{pmatrix}\begin{pmatrix}
        h\big(1 - \cos f\big) & -h\sin f\\
        -h\sin f & h\big(1+\cos f\big)
    \end{pmatrix}\begin{pmatrix}
        u_1 \\
        u_2
    \end{pmatrix}\right]\nonumber\\
    &= 2h\int_0^\infty r\mathrm{d}r\,\bigg(u_2 \cos (f/2) - u_1 \sin (f/2)\bigg)^2\nonumber\\
    &\geq 0\;.
\end{align}
We can then conclude that
\begin{equation}
    \varepsilon \geq m_-\;,
\end{equation}
so the lowest energy state for $e=0$ in the Skyrmion background sits at the threshold $\varepsilon_{\mathrm{min,NR}} = m_-$ for scattering states and does not produce a bound state. It is possible that $\mathcal{O}(h/m)$ corrections or relativistic corrections that we have dropped can generate a potential leading to the formation of a bound state, but numerically we do not find any evidence for such states.

\section{Numerical results}
\label{sec:num}

\subsection{Numerical methods}
\begin{figure}[t]
    \centering
    \begin{subfigure}[b]{0.49\textwidth}
    \centering
    \includegraphics[width=\linewidth]{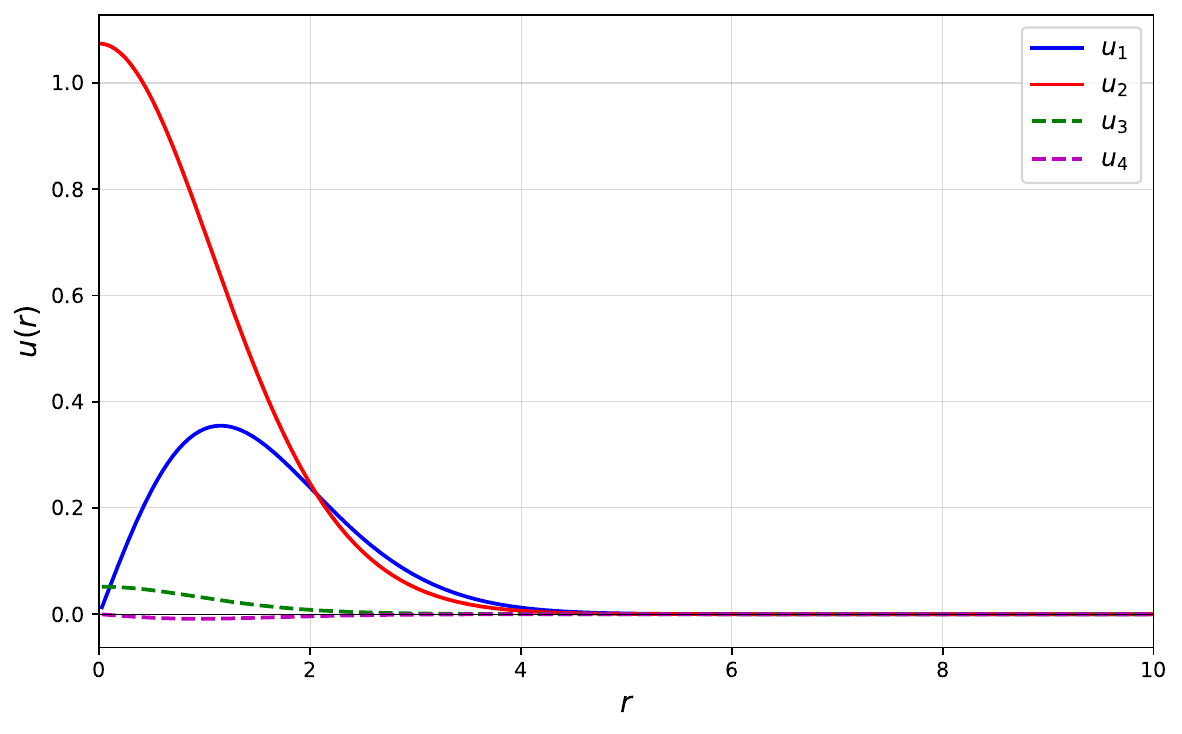}
    \vskip-2ex
    \caption{\small $l=0$}
    \label{fig:wavefn_l0}
    \end{subfigure}
    \hfill
    \begin{subfigure}[b]{0.49\textwidth}
    \centering
    \includegraphics[width=\linewidth]{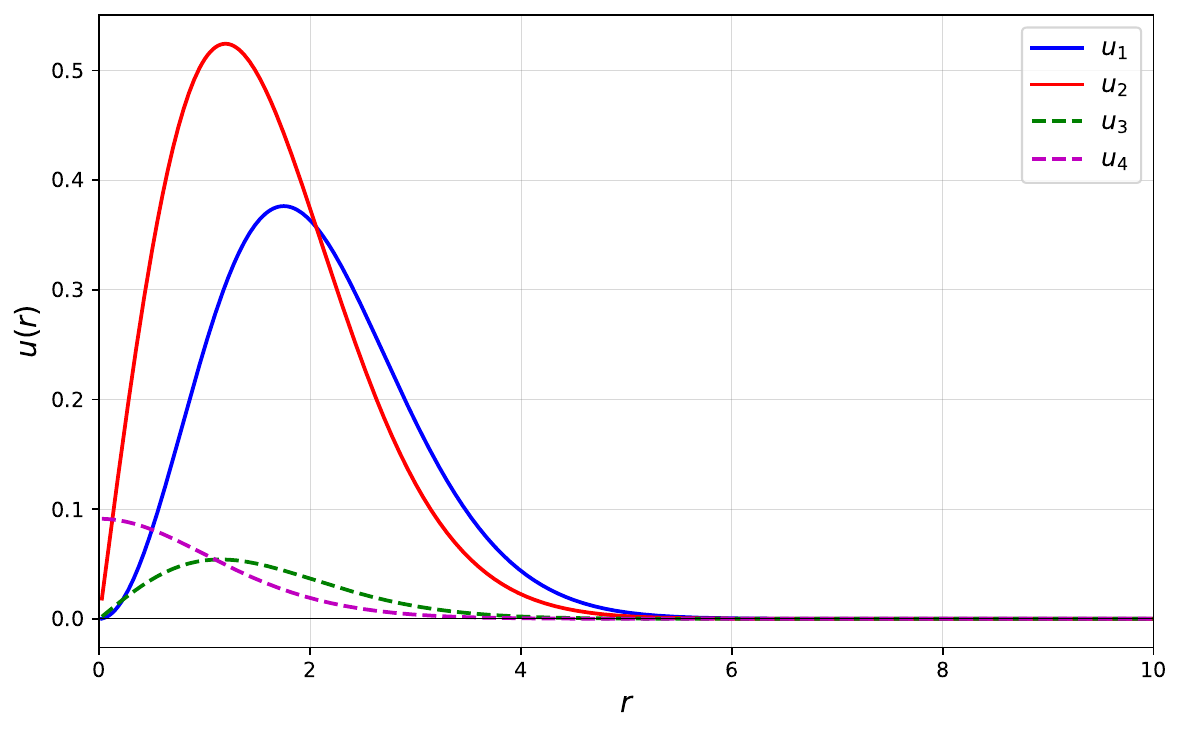}
    \vskip-2ex
    \caption{\small $l=-1$}
    \label{fig:wavefn_lm1}
    \end{subfigure}
    \hfill
    \begin{subfigure}[b]{0.49\textwidth}
    \centering
    \includegraphics[width=\linewidth]{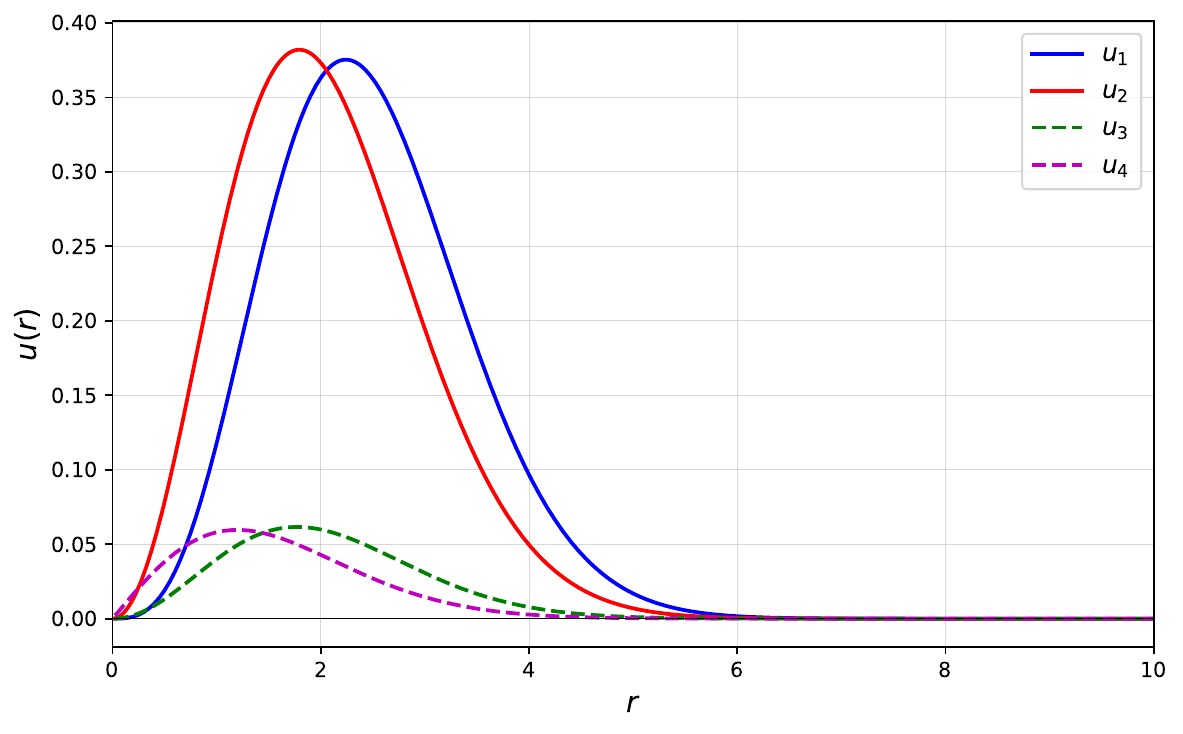}
    \vskip-2ex
    \caption{\small $l=-2$}
    \label{fig:wavefn_lm2}
    \end{subfigure}
    \hfill
    \begin{subfigure}[b]{0.49\textwidth}
    \centering
    \includegraphics[width=\linewidth]{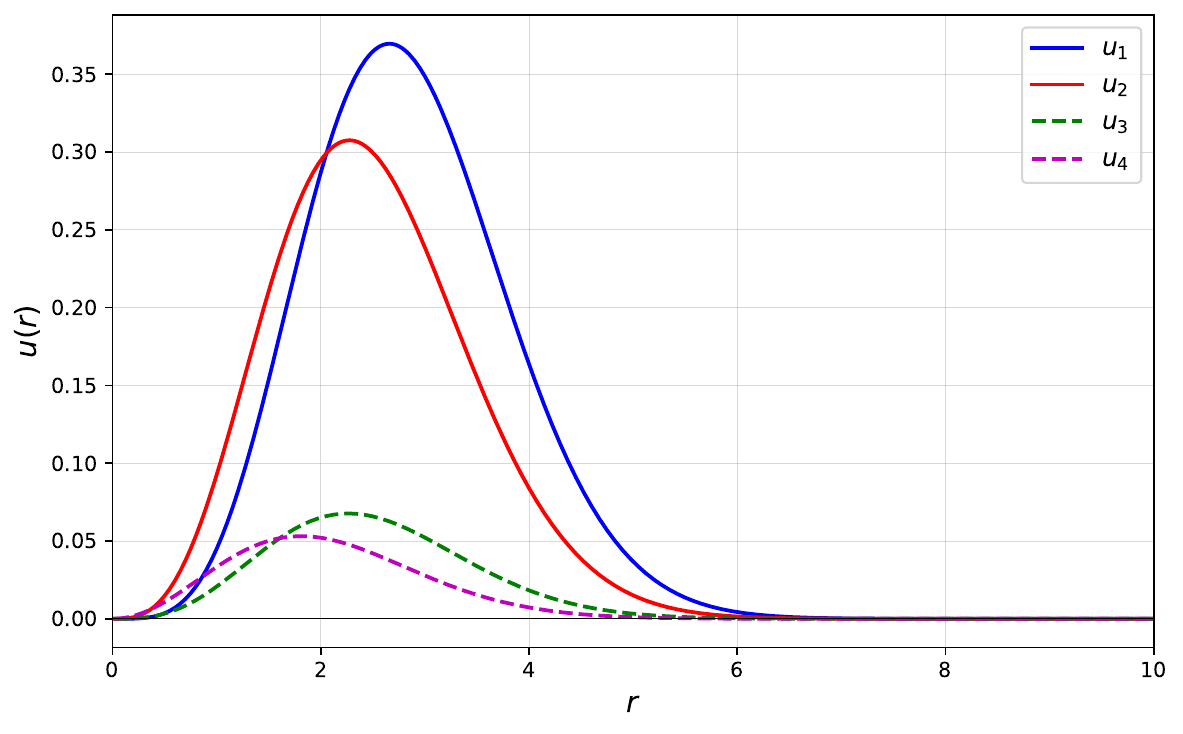}
    \vskip-2ex
    \caption{\small $l=-3$}
    \label{fig:wavefn_lm3}
    \end{subfigure}
 
    \caption{\small Bound state wavefunctions for $l=0,-1,-2,-3$ with $e=1$, $m=10$, and at $h=2$. All states are normalized to $Q=+1$.}
    \label{fig:wavefn_bound_charge}
\end{figure}

We solve the full four-component Dirac equation \cref{eq:bddbck} numerically by discretizing the system on a grid. First, the Skyrmion profile ansatz \cref{eq:santz} satisfying the equations of motion derived from the Lagrangian \cref{eq:Lskym} is solved numerically using \texttt{scipy.integrate.solve\_bvp} \cite{2020SciPy-NMeth}. The Hamiltonian \cref{eq:fermHam} is discreteized on the same grid with derivatives approximated using finite difference methods. The resulting sparse matrix can be diagonalized efficiently using \texttt{scipy.sparese.linalg.eigs}. The resulting eigenvectors are normalized such that $Q = +1$ such that we focus on states with total particle charge $1$ where a bound state corresponds to a localized wavefunction with negative energy. Generally, due to fermion doubling on the lattice there will be two eigenvectors with nearly identical energy eigenvalues \cite{Nielsen:1980rz}. We remove unphysical solutions which arise due to fermion doubling on the lattice by flagging which of the two oscillates on the lattice scale, and rejecting it as a physical solution. The surviving wavefunctions are smooth and exhibit the localized behavior expected of bound states, see for example the wavefunctions of \cref{fig:wavefn_bound_charge} In the non-relativistic limit, the two equations for $u_1$ and $u_2$ \cref{eq:NR_eqns} are solved in an identical manner to the relativistic case by discretizing the Hamiltonian and diagonalizing the resulting sparse matrix.

\subsection{Bound states of fermions in magnetic Skyrmion background at the critical coupling}
\label{sec:bd_charged}
\begin{figure}[t]
    \centering
    \includegraphics[width=1\linewidth]{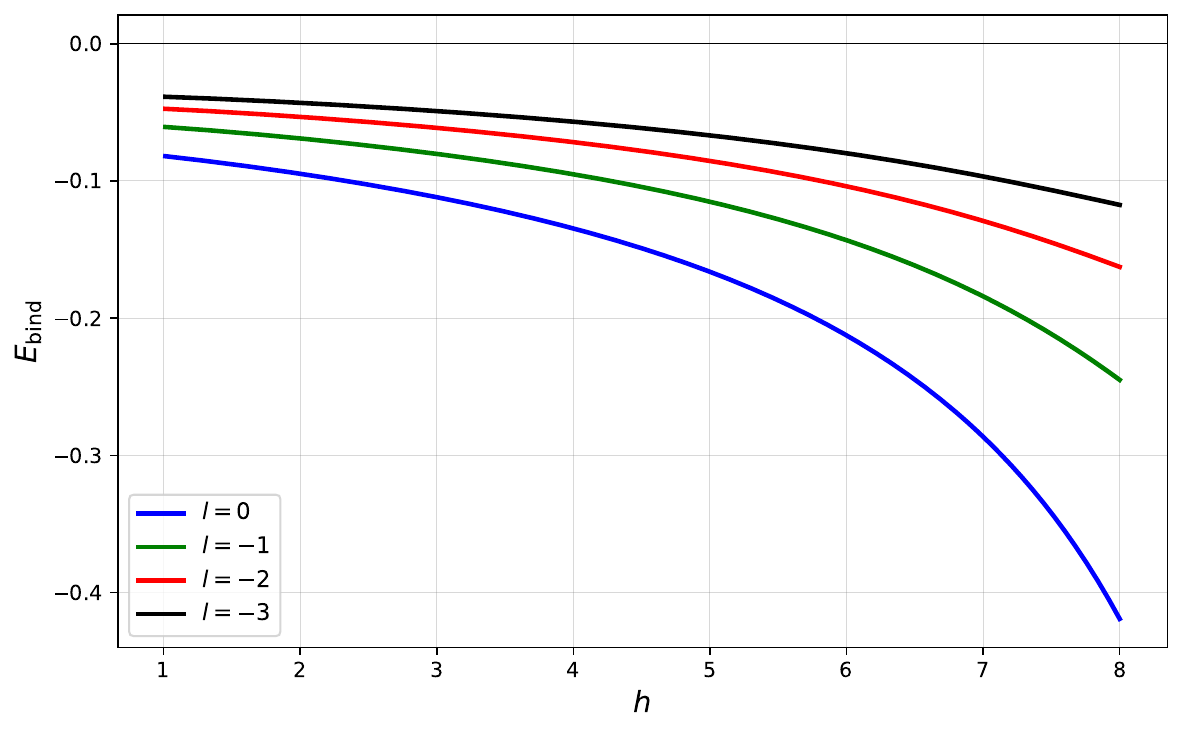}
    \vskip-2ex
    \caption{\small Binding energy $E_{\mathrm{bind}}=\varepsilon-\varepsilon_{\mathrm{min}}$ for fermions with mass $m=10$, angular momentum $l = 0,-1,-2,-3,$ and electric charge $e=1$ in the magnetic Skyrmion background at critical coupling. No bound states are observed outside of $e>0$ with $l \leq 0$ consistent with the non-relativistic analysis, and in this regime only one bound state is found per $l$.}
    \label{fig:bound_m10_e1}
\end{figure}

In this section, we investigate bound states of fermions in the magnetic Skyrmion background at critical coupling $(\lambda,\kappa,\mu) = (0,1,1)$. Whether a given state is bound to the Skyrmion is determined by comparing the energy of the fermion in the Skyrmion background to the ground state energy of the fermion in the absence of the Skyrmion (i.e. set $f(r) = 0$). As discussed in \cref{sec:ENoSkyrm}, the ground state energy in the absence of the Skyrmion is given by the standard relativistic 2D Landau levels given by \cref{eq:E_Landau}. We further enforce that such a state is a particle rather than antiparticle by normalizing wavefunctions to $Q=+1$. We work in the non-relativistic regime $h \ll m$ where the quantum mechanical calculations can be trusted and find no evidence of bound states for neutral fermions, and bound states only for $e > 0$ and $l \leq 0$. In the regime $e>0$ and $l \leq 0$, we observe only a single state per $l$ bound to the Skyrmion. The binding energy $E_{\mathrm{bind}}=\varepsilon-\varepsilon_{\mathrm{min}}$ as a function of the fermion-Skyrmion coupling $h$ for $m=10$ and $e=1$ is provided in \cref{fig:bound_m10_e1}.

As $|l|$ increases, the fermion generically has larger kinetic energy and is less tightly bound. This can also be understood as the strengthening of a centrifugal barrier in the effective potential leading to the fermion-Skyrmion bound state. We find numerically that the best fit for $E_{\mathrm{bind}}$ as a function of $l$ is a shifted power law, with
\begin{equation}
    E_{\mathrm{bind}} = -\frac{A}{(|l|+c)^\alpha}\;.
\end{equation}

\begin{figure}[t]
    \centering
    \includegraphics[width=1\linewidth]{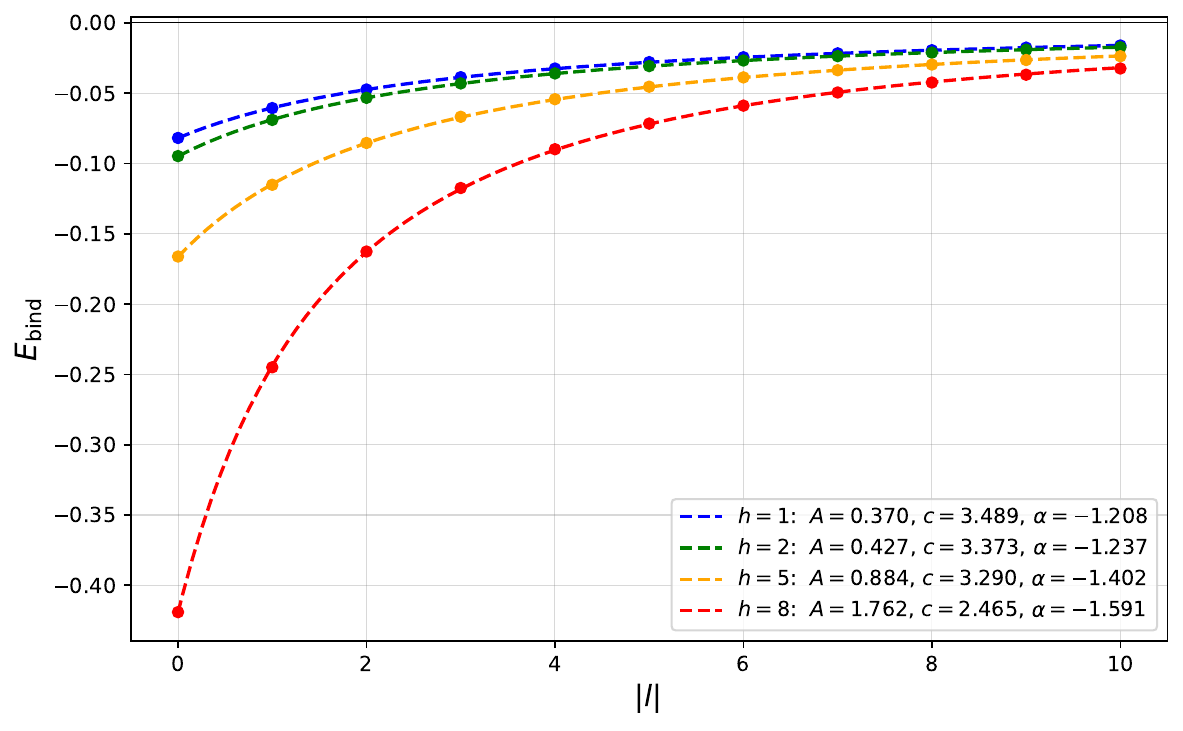}
    \vskip-2ex
    \caption{\small Numerical data (solid dots) for the binding energy for the magnetic Skyrmion background with $m=10$, $e=1$, $l \leq 0$, for $h=1,2,5,8$. The best fit for the shifted power law $E_{\mathrm{bind}} = -A/(|l|+c)^\alpha$ are given as the dashed colored line, with the parameters $A$, $c$, and $\alpha$ for each fit provided in the legend.}
    \label{fig:binding_vs_l}
\end{figure}

\subsection{Bound states of fermions in the mixing background}

\begin{figure}[t]
    \centering
    \begin{subfigure}[b]{0.49\textwidth}
    \centering
    \includegraphics[width=\linewidth]{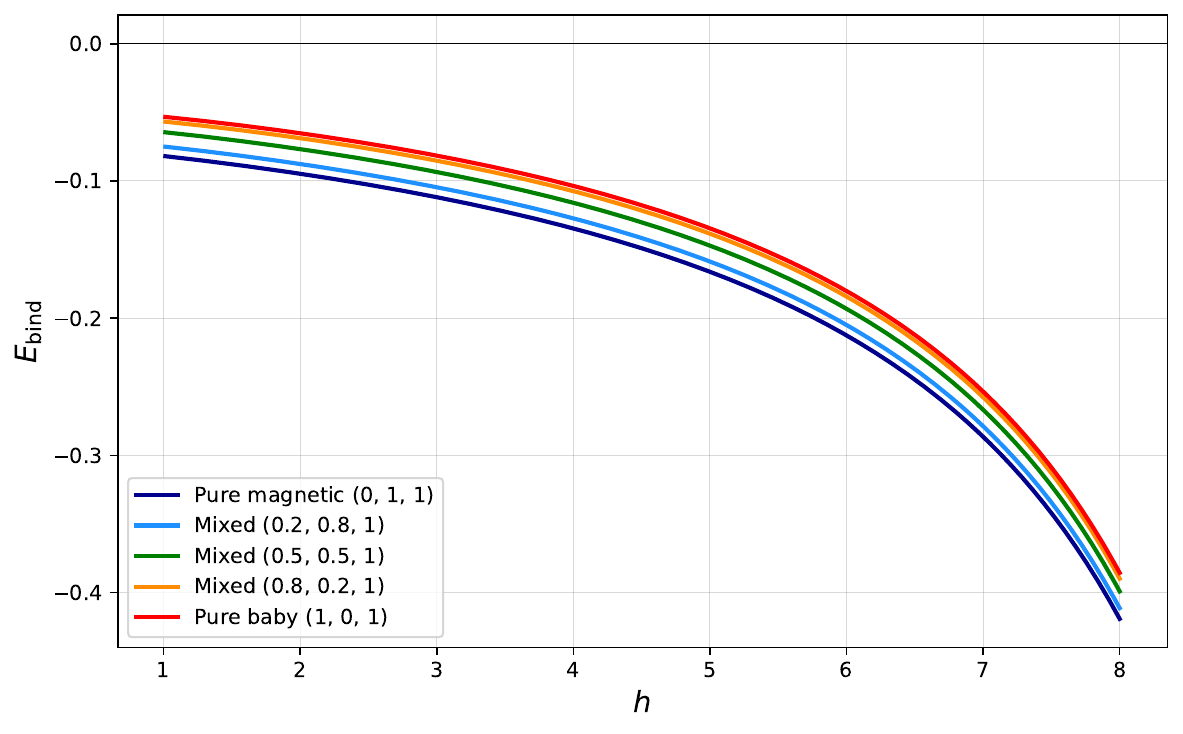}
    \vskip-2ex
    \caption{\small $l=0$}
    \label{fig:mixed_l0}
    \end{subfigure}
    \hfill
    \begin{subfigure}[b]{0.49\textwidth}
    \centering
    \includegraphics[width=\linewidth]{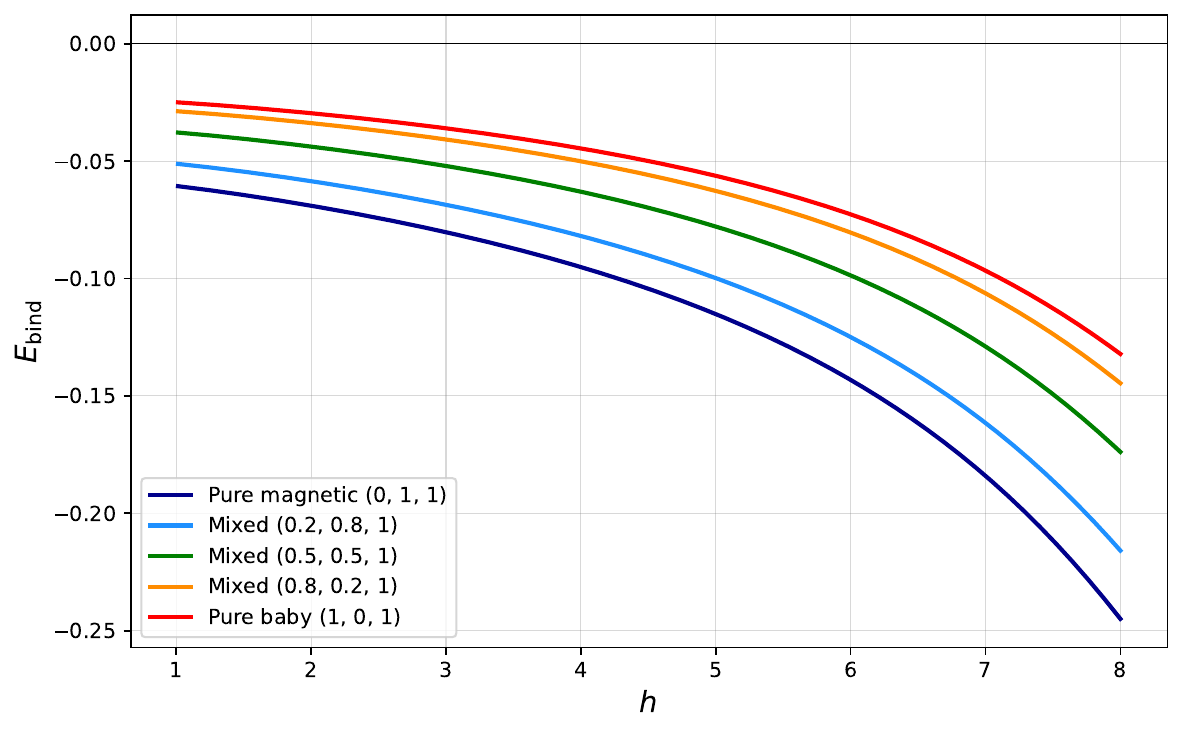}
    \vskip-2ex
    \caption{\small $l=-1$}
    \label{fig:mixed_lm1}
    \end{subfigure}
    \hfill
    \begin{subfigure}[b]{0.49\textwidth}
    \centering
    \includegraphics[width=\linewidth]{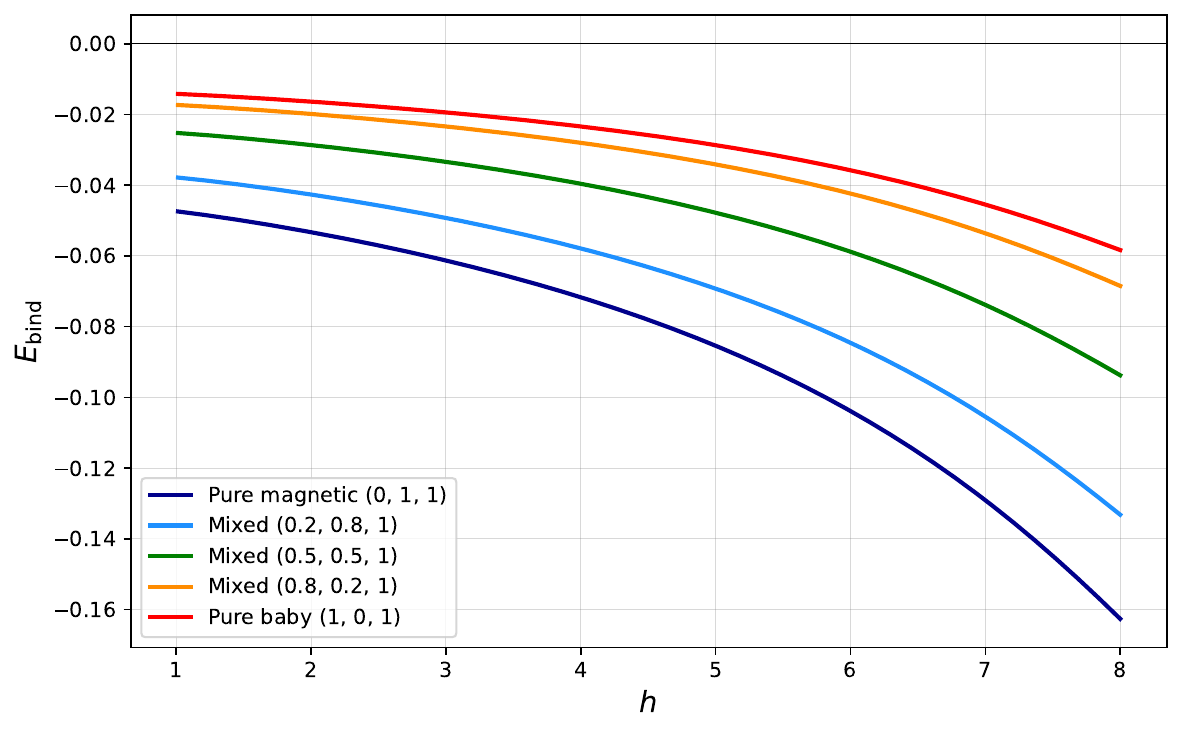}
    \vskip-2ex
    \caption{\small $l=-2$}
    \label{fig:mixed_lm2}
    \end{subfigure}
    \hfill
    \begin{subfigure}[b]{0.49\textwidth}
    \centering
    \includegraphics[width=\linewidth]{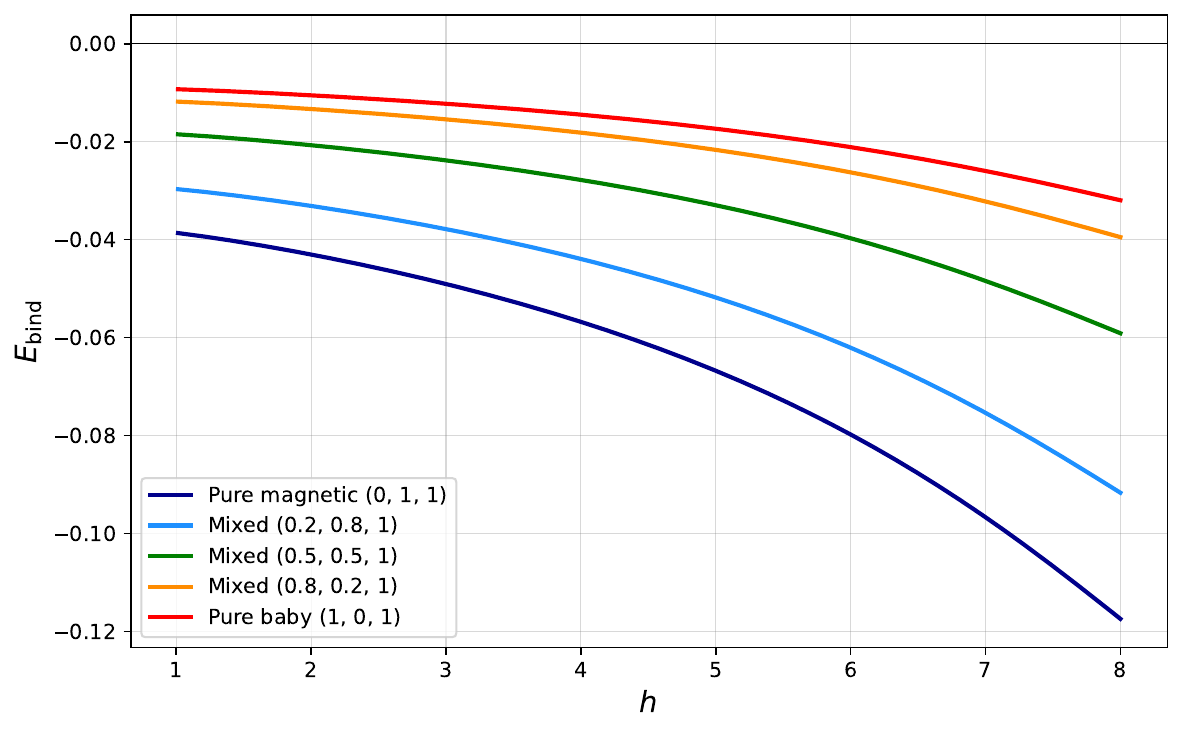}
    \vskip-2ex
    \caption{\small $l=-3$}
    \label{fig:mixed_lm3}
    \end{subfigure}
 
    \caption{\small Binding energy in the mixed baby and magnetic Skyrmion background for $e = 1$, $l \leq 0$ with $h < m$, where $m = 10$. The magnetic Skyrmion background provides the deepest binding, while the baby Skyrmion provides the weakest binding.}
    \label{fig:mixed_bound}
\end{figure}

One can also consider the fermion in a mixed background of the magnetic and baby Skyrmion. The arguments of \cref{ssec:analytic_NR_e_neq_0} still carry through in this case, and numerically we again find evidence for bound states in the mixing (as well as pure baby) Skyrmion backgrounds only when $e>0$ and $l \leq 0$ for $h < m$. Numerical results for the binding with $e = 1$ and $l = 0,-1,-2,-3$ with $m=10$ are provided in \cref{fig:mixed_bound}.

A hierarchy in binding strength emerges across all angular momenta. The pure magnetic Skyrmion background $(\lambda,\kappa)=(0,1)$ provides the deepest binding, the pure baby Skyrmion $(\lambda,\kappa)=(1,0)$ the shallowest, and the mixed backgrounds interpolate monotonically between these two extremes. 
Increasing the baby Skyrmion fraction $\lambda$ at fixed $\mu=1$ consistently weakens the fermion-Skyrmion binding, reflecting the difference in the spatial profiles of the two Skyrmion types and the resulting depth of the effective potential experienced by the fermion. For sufficiently large $h$, the binding increases monotonically in magnitude with the coupling $h$ across all backgrounds.

This can be understood from a simple estimation. As seen in \cref{fig:skypflh0}, the baby Skyrmion profile is more spatially concentrated than the magnetic Skyrmion profile. The effective potential experienced by the fermion can then be approximated as a finite spherical well of depth $V_0$ and width $a$, and in the regime where the well is wide compared to the fermion size the binding energy takes the form
\begin{align}
    E_{\rm bind} \sim -V_{0} + \frac{g(l^2)}{a^2},
\end{align}
where $g(l^2) > 0$ is a coefficient that depends on the angular momentum. Since the magnetic Skyrmion profile has a larger spatial extent than the baby Skyrmion, the $g(l^2)/a^2$ contribution is smaller, resulting in deeper binding. Conversely, the more concentrated baby Skyrmion profile leads to a larger centrifugal energy cost and hence weaker binding, with the mixed backgrounds interpolating between the two extremes. This is consistent with the results shown in \cref{fig:mixed_bound}.

The effect of angular momentum on the relative spread between backgrounds is also notable. For $l=0$, the binding energies for all five configurations are comparatively close to one another throughout the range of $h$ shown, and all curves track together as $h$ increases. 
For larger $|l|$, the gap in binding between the pure magnetic and pure baby backgrounds grows more significantly with increasing $h$, while the two are more comparable at small $h$. This behavior is consistent with the picture from \cref{sec:bd_charged} in which the centrifugal barrier suppresses binding at large $|l|$, since the stronger effective potential of the magnetic Skyrmion background is better able to overcome this barrier. 
These results suggest that tuning the Skyrmion background composition offers a continuous handle on the binding energy of the fermion-Skyrmion system, with the magnetic Skyrmion providing the most favorable conditions for tightly bound states.

\subsection{Comparison to non-relativistic limit}
\begin{figure}[t]
    \centering
    \begin{subfigure}[b]{0.49\textwidth}
    \centering
    \includegraphics[width=\linewidth]{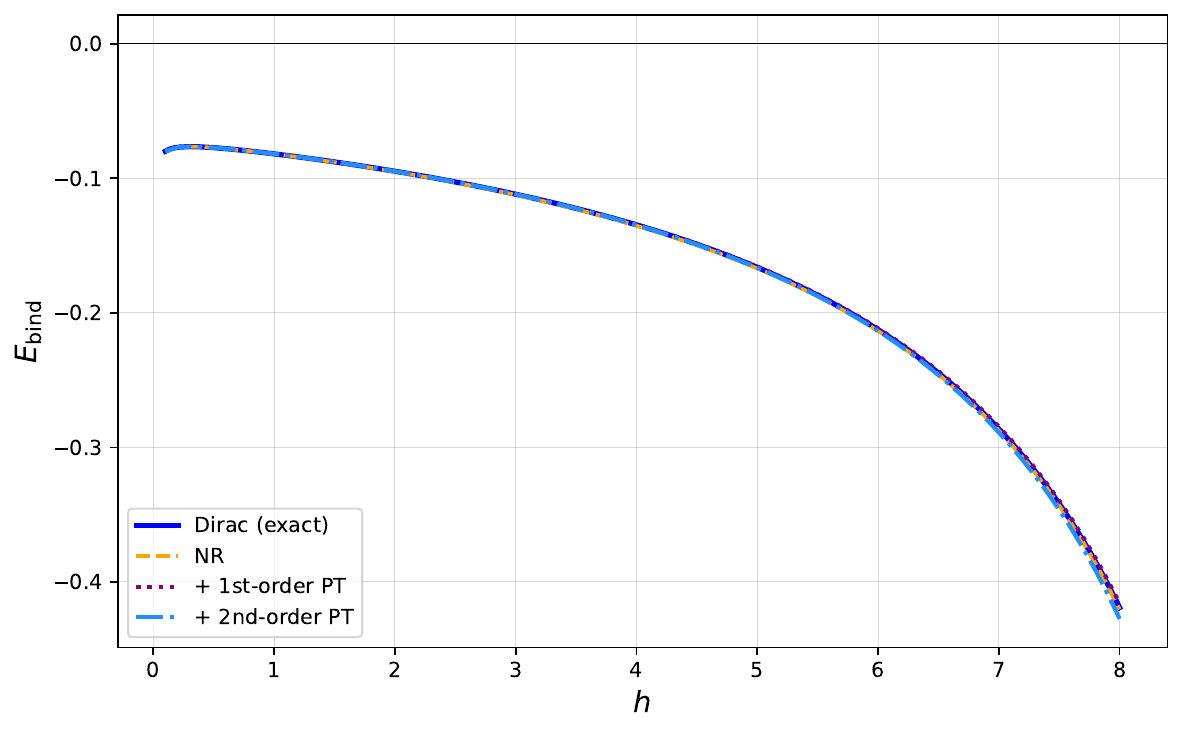}
    \vskip-2ex
    \caption{\small $l=0$}
    \label{fig:NR_l0}
    \end{subfigure}
    \hfill
    \begin{subfigure}[b]{0.49\textwidth}
    \centering
    \includegraphics[width=\linewidth]{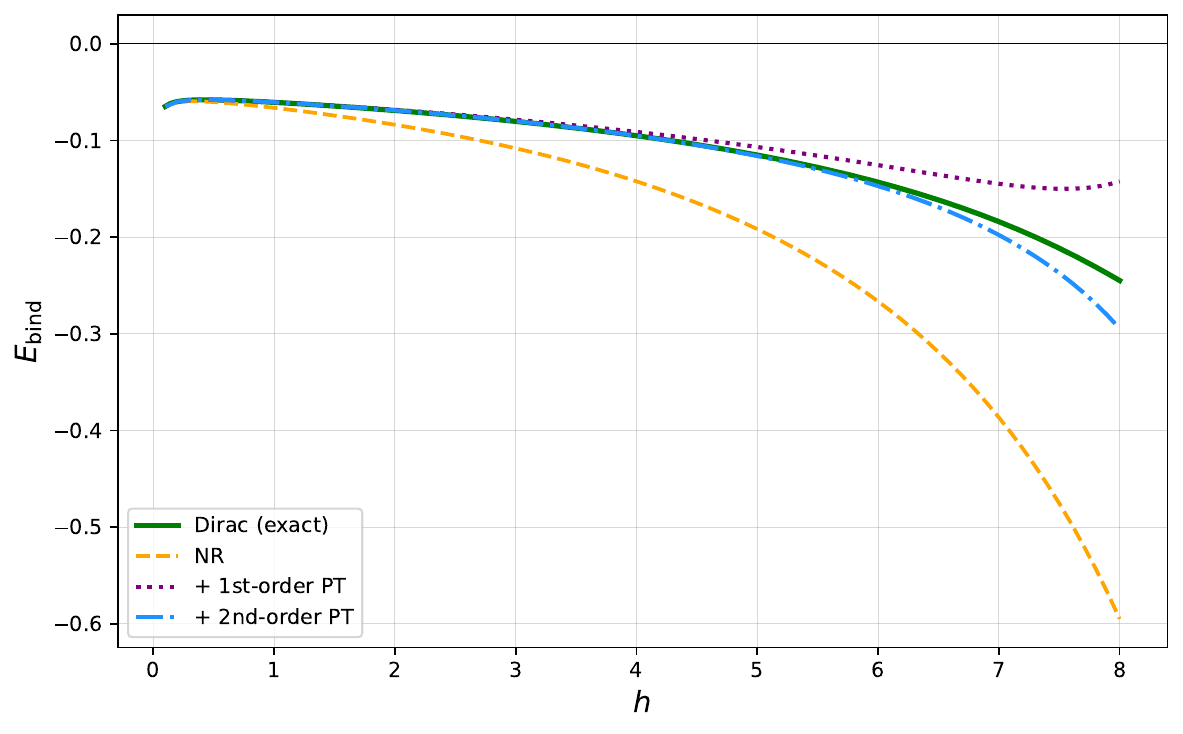}
    \vskip-2ex
    \caption{\small $l=-1$}
    \label{fig:NR_lm1}
    \end{subfigure}
    \hfill
    \begin{subfigure}[b]{0.49\textwidth}
    \centering
    \includegraphics[width=\linewidth]{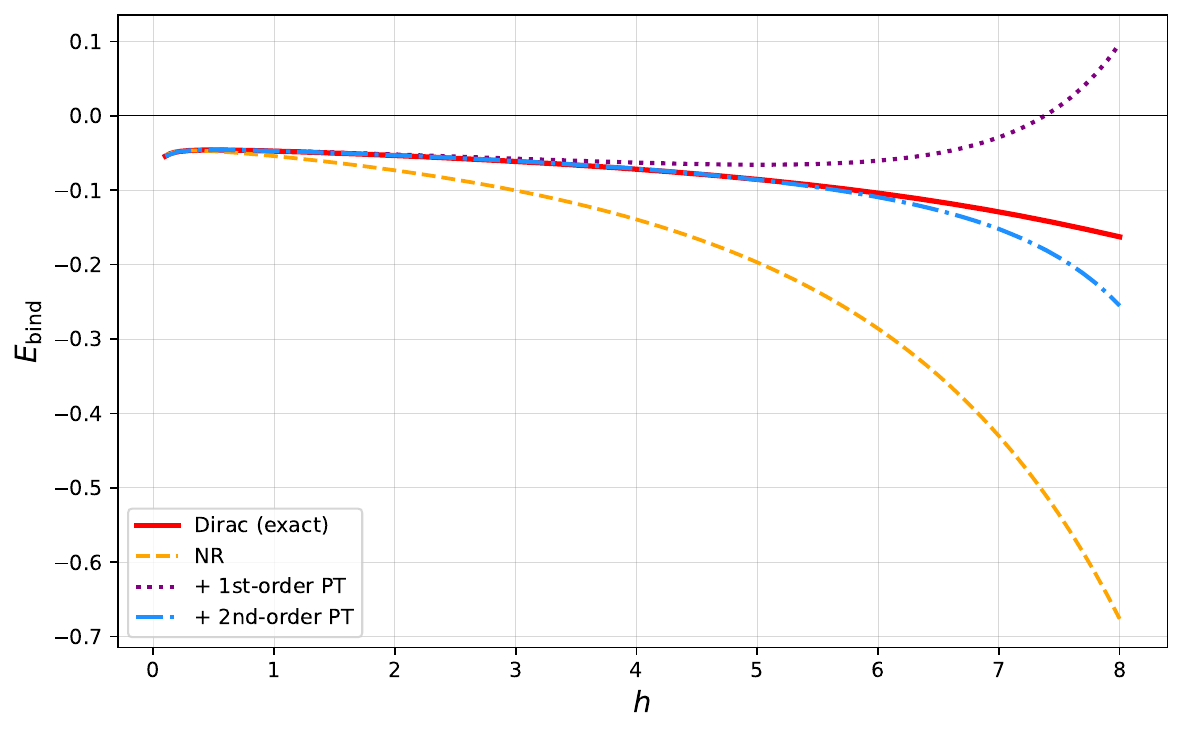}
    \vskip-2ex
    \caption{\small $l=-2$}
    \label{fig:NR_lm2}
    \end{subfigure}
    \hfill
    \begin{subfigure}[b]{0.49\textwidth}
    \centering
    \includegraphics[width=\linewidth]{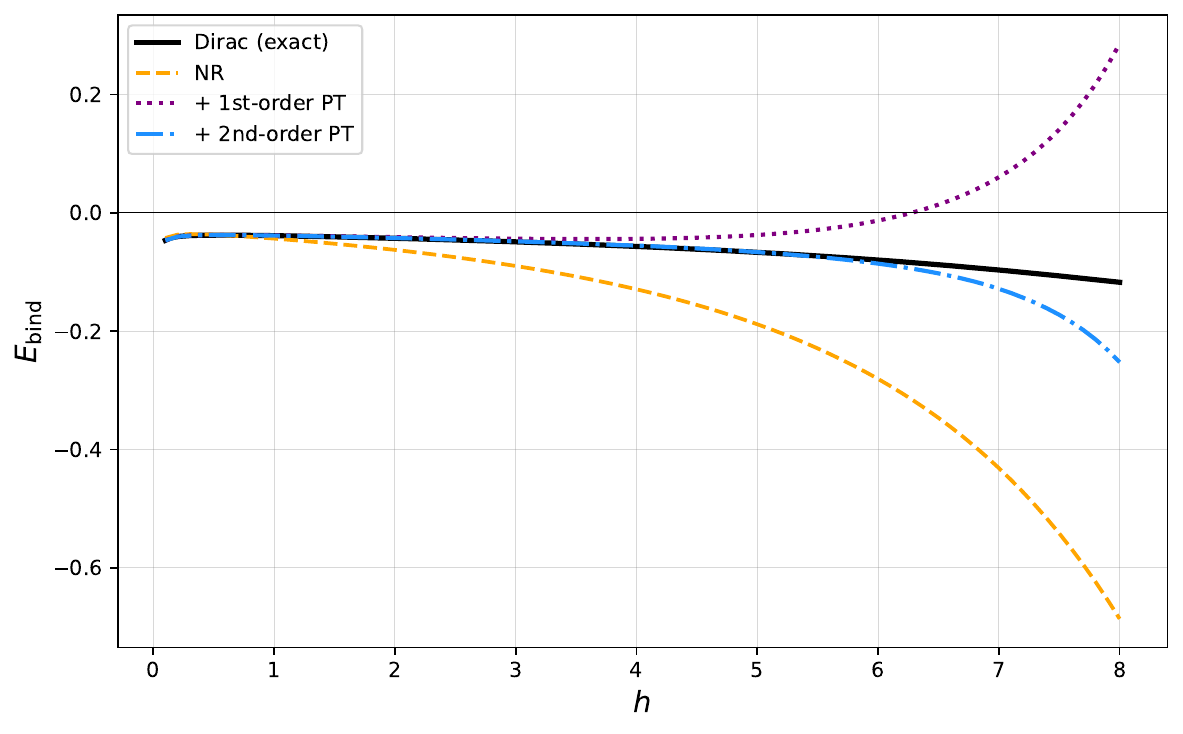}
    \vskip-2ex
    \caption{\small $l=-3$}
    \label{fig:NR_lm3}
    \end{subfigure}
 
    \caption{\small Comparison of the binding energy as computed by the fully relativistic Dirac equations, NR equations, and higher order $h/m$ corrections to the NR system.}
    \label{fig:NR_compare}
\end{figure}

Here we compare binding energy predicted by the NR system described by \cref{eq:NR_eqns} to the binding energy computed from the fully relativistic system. The NR system from \cref{eq:NR_eqns} can be solved in the same way as the full four-component Dirac equations by discretizing the Hamiltonian and diagonalizing the resulting sparse matrix using \texttt{scipy.sparse.linalg.eigs}. The binding energy computed by the NR system to lowest order consistently predicts a tighter binding than the fully relativistic system  $l \neq 0$ as the NR system systematically underestimates the centrifugal barrier that reduces the binding energy. We can include higher order $h/m$ corrections to the NR system by inverting the first two equations of \cref{eq:bddbck} for $u_3$ and $u_4$, and obtaining a set of second order differential equations for $u_1$ and $u_2$ which retain the explicit $\varepsilon$ dependence rather than applying the NR limit which replaces $\varepsilon$ with the appropriate $m_{\pm}$ rest mass for the channel. The resulting equations can be written in the form
\begin{equation}
    \varepsilon \begin{pmatrix}
        u_1\\
        u_2
    \end{pmatrix} = H_{\mathrm{exact}}(\varepsilon) \begin{pmatrix}
        u_1\\
        u_2
    \end{pmatrix}\;.
\end{equation}
The difference between $H_{\mathrm{exact}}(\varepsilon)$ and the $H_{\mathrm{NR}}$ as given in \cref{eq:NR_eqns} can be treated as a perturbation to the NR system,
\begin{equation}
    V_{\mathrm{NR,pert}}(\varepsilon) = H_{\mathrm{exact}}(\varepsilon) - H_{\mathrm{NR}}\;.
\end{equation}
Solving the NR system for the bound state with energy $\varepsilon_0$ and wavefunction $\psi$, we can compute higher-order corrections to the binding energy through standard perturbation theory, such as the first order correction $(\varepsilon_0)^{(1)} = \bra{\psi}V_{\mathrm{NR,pert}}(\epsilon_0)\ket{\psi}$. Notice that we evaluate the perturbation $V_{\mathrm{NR,pert}}(\varepsilon)$ at the leading order energy $\varepsilon_0$ computed from the NR system. Including first- and second-order corrections to the binding energy, the predicted binding energy matches the fully relativistic computation to the percent level for $h\lesssim 6$ when $m=10$, as shown in \cref{fig:NR_compare}. We have also verified that the corrections to the binding energy reduce the discrepancy between the fully relativistic and NR predictions by a factor of $h/m$ for each successive correction included in the NR computation, as expected for the NR expansion.

\section{Discussion and Conclusion \label{sec:cnclsn}}

In this paper we have studied a single Dirac fermion coupled to a $(2+1)$-dimensional Skyrme model that interpolates between the magnetic Skyrmion at the critical coupling, $(\kappa,\lambda,\mu)=(1,0,1)$, and the baby Skyrmion, $(\kappa,\lambda,\mu)=(0,1,1)$,with the fermion coupled both to the Skyrmion isospin through the term $h\,\vec{S}\cdot\bar{\psi}\vec{\tau}\psi$ and, through its electric charge $e$, to the uniform background field $\vec{B}=\mu^{2}\hat{\vec{z}}$ associated with the quadratic Zeeman potential. 
We have characterized the fermion bound-state spectrum across angular momentum $l$, mass $m$, charge $e$ and coupling $h$, both analytically in the non-relativistic limit and numerically using the full Dirac equation, and the variational method for the quantum mechanical reduction.

Our analytic results of \cref{sec:NR_limit} provides a criterion for the existence of bound states. In the absence of the Skyrmion background, the spectrum reduces to two shifted families of Landau levels with minimal energies $A_1$ and $A_2$, and a variational argument shows that a bound state can only form when $A_1>A_2$, which translates into $|e|(|l-1|-|l|)+e>0$ and singles out the sector $e>0$, $l\leq 0$. 
The numerical results of Sec.~\ref{sec:bd_charged} confirm this prediction in the magnetic Skyrmion background and show that the rule persists numerically up to $h/m\sim\mathcal{O}(1)$ as long as $h<m$, indicating that the obstruction is robust beyond the strict non-relativistic expansion.

Within the allowed sector we observe exactly one bound state per angular momentum,
whose binding energy decreases monotonically with $h$, encoding the weaker binding as the centrifugal barrier as $|l|$ grows. 
Extending the analysis to the mixed Skyrme background, see \cref{fig:mixed_bound}, reveals a hierarchy. The binding between a fermion and the magnetic Skyrmion at the critical coupling is the strongest one, and can be interpolated to that of the baby Skyrmion which has the weakest binding among the cases we take into account.
This can be inferred from the finite-well estimate related to the spatial profiles in \cref{fig:skypflh0}
The more diffuse magnetic-Skyrmion profile minimizes the centrifugal cost while the more compact baby-Skyrmion profile strengthens it.
Also, in these Skyrmion backgrounds the gap between curves is small at $l=0$ and increase with $|l|$. Comparing between all these cases, we can think of that tuning the ratio $\lambda/\kappa$ with their sum fixed is thus viewed as an interpolation on the fermion--Skyrmion binding energy.

\begin{figure}[t]
    \centering
    \begin{subfigure}[b]{0.49\textwidth}
    \centering
    \includegraphics[width=\linewidth]{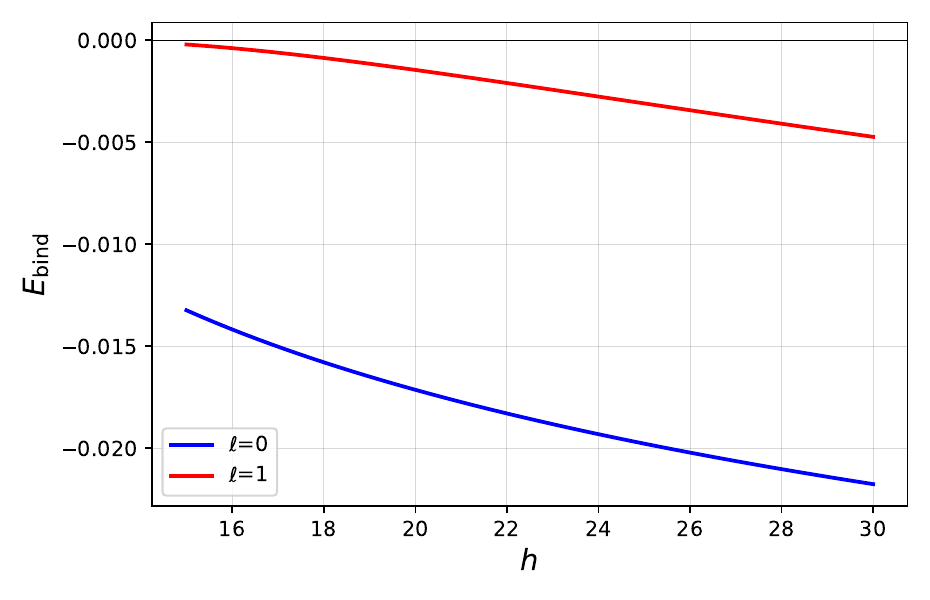}
    \vskip-2ex
    \caption{\small $e=0$}
    \label{fig:hgme0}
    \end{subfigure}
    \hfill
    \begin{subfigure}[b]{0.49\textwidth}
    \centering
    \includegraphics[width=\linewidth]{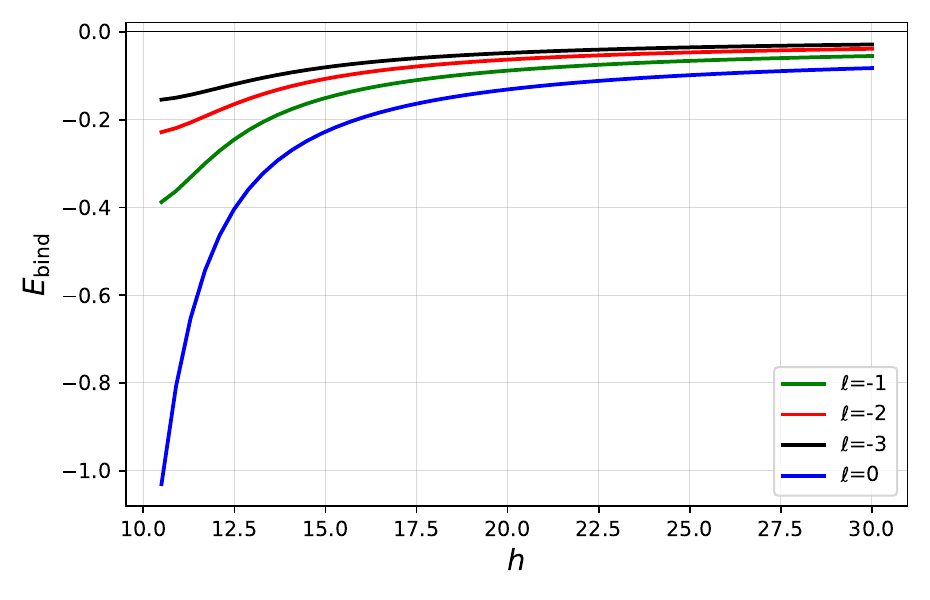}
    \vskip-2ex
    \caption{\small $e=1$}
    \label{fig:hgme1}
    \end{subfigure}
    \caption{\small Binding energy for fermions with mass $m=10$ in the Skyrmion background of the critical coupling with the coupling constant $h$ greater than $m$.}
    \label{fig:hgm}
\end{figure}
In addition, we restrict our attention throughout this paper to the parameter region $h<m$, in contrast to previous studies \cite{Perapechka:2018yux,Perapechka:2019upv,Barsanti:2021vhd} considering $h>m$. 
Our numerical scheme is nonetheless not limited to this window. For $h>m$, applying the same procedure outlined in \cref{sec:num}, one continues to find bound states analogous to those reported in the aforementioned references. See for example, \cref{fig:hgm}. 
In the neutral case, only fermions with $l=0,1$ admit bound states, whereas for a charged fermion several Landau levels (labeled by different $l$) become bound to the Skyrmion background. 
A conceptual subtlety, however, may arise from this enlarged spectrum. Once $h$ exceeds $m$, the potential well generated by the Skyrmion can become sufficiently deep that the bound level dives toward the negative-energy continuum, where particle production can happen. The numerically obtained state can then no longer be interpreted as a genuine single-particle bound state. 
We leave a detailed investigation of this regime to future work.

Lastly, those bound states localized to a Skyrmion in the present study may have direct experimental implications. When the trapped fermion carries non-zero electric charge, the composite behaves as a localized charge carrier dressed by the topological texture, and its transport response should deviate from that of a bare Skyrmion. A natural way to expose this difference is via Hall-effect measurements, in close analogy with the classic studies of carrier transport in disordered and hopping systems \cite{FRHall,Friedman01081978,Friedman01101981,BUTCHER198089,BMovaghar_1981}, see
\cite{shklovskii2013electronic} for a comprehensive review. 
The complementary spectroscopic signature should be accessible through scattering measurements \cite{PhysRevResearch.2.013247,Loginov:2021rka,Loginov:2024nmi}. 
Some natural extensions of this work include a quantitative computation of the modified Hall conductivity and the generalization to Skyrmion lattices where fermion trapping competes with band formation, which we leave for future study.

\section*{Acknowledgements}
This work was supported in part by the US National Science Foundation under grant PHY-2210283. A.S. was supported by the National Science Foundation Graduate Research Fellowship Program. This material is based upon work supported by the National Science Foundation Graduate Research Fellowship Program under Grant No. DGE-2235784. Any opinions, findings, and conclusions or recommendations expressed in this material are those of the authors and do not necessarily reflect the views of the National Science Foundation.

\bibliographystyle{apsrev4-1}
\bibliography{draft.bbl}

\end{document}